\documentclass[a4paper,10pt,twoside]{cpc-hepnp}

\usepackage{multicol}
\usepackage{graphicx}
\usepackage{booktabs}
\usepackage{amssymb,bm,mathrsfs,bbm,amscd}
\usepackage[tbtags]{amsmath}
\usepackage{lastpage}

\newcommand{\psp}{\psi^{\prime}}

\newcommand{\pspp}{\psi^{\prime \prime}}

\newcommand{\jpsi}{J/\psi}
\newcommand{\EE}{e^+e^-}

\newcommand{\pp}{\pi^+\pi^-}
\newcommand{\kk}{K^+K^-}
\newcommand{\kskl}{K^0_SK^0_L}

\newcommand{\VP}{1^-0^-}
\newcommand{\PP}{0^-0^-}
\def\eref#1{(\ref{#1})}
\def\Journal#1#2#3#4#5{{#5}. {#1}, {#4}, {\bf #2}: #3}

\def\IJMPA{Int. J. Mod. Phys. A}

\def\NIMA{Nucl. Instrum. Methods A}

\def\NPB{Nucl. Phys. B}

\def\PLB{Phys. Lett. B}

\def\PRL{Phys. Rev. Lett.}
\def\PRD{Phys. Rev. D}

\def\CPC{China Physic C}

\begin{document}

%\fancyhead[co]{\footnotesize WANG Bo-Qun et al: Fast computation of observed
%cross section for $\psp \to PP$ decays}

\footnotetext{This paper will be published in Chinese Physics C}

\title{Fast computation of observed cross section \\
for $\psp \to PP$ decays\thanks{Supported by National Natural
Science Foundation of China (10491303, 10775412, 10825524, 10835001, 10935008),
the Instrument Developing Project of the Chinese Academy of Sciences
(YZ200713), Major State Basic Research Development Program (2009CB825200,
2009CB825206), and Knowledge Innovation Project of The Chinese Academy of
Sciences (KJCX2-YW-N29).} }

\author{%
		WANG Bo-Qun$^{1,2;1)}$\email{wangbq@ihep.ac.cn}%
\quad	MO Xiao-Hu$^{2}$%
\quad   WANG Ping$^{2}$%
\quad   BAN Yong$^{1}$
}

\maketitle

\address{%
$^1$ School of Physics, Peking University, Beijing 100871, China\\
$^2$ Institute of High Energy Physics, Chinese Academy of Sciences, Beijing
100049, China\\
}

\begin{abstract}
It has been conjectured that the relative phase between strong and
electromagnetic amplitudes is universally $-90^{\circ}$ in charmonium decays.
$\psp$ decaying into pseudoscalar pair provides a possibility to test this
conjecture.  However, the experimentally observed cross section for such a
process is depicted by the two-fold integral which takes into account the
initial state radiative (ISR) correction and energy spread effect. Using the
generalized linear regression approach, a complex energy-dependent factor is
approximated by a linear function of energy. Taking advantage of  this
simplification, the integration of ISR correction can be performed and an
analytical expression with accuracy at the level of 1\% is obtained. Then, the
original two-fold integral is simplified into a one-fold integral, which
reduces the total computing time by two orders of magnitude. Such a simplified
expression for the observed cross section usually plays an indispensable role
in the optimization of scan data taking, the determination of systematic
uncertainty, and the analysis of data correlation.
\end{abstract}



\begin{multicols}{2}

\section{Introduction}

The relative phase between the strong and the electromagnetic amplitudes of the
charmonium decays is a basic parameter in understanding decay dynamics. Studies
have been carried out for many $\jpsi$ two-body decay modes:
$1^-0^-$~\cite{dm2exp,mk3exp}, $0^-0^-$~\cite{a00,lopez,a11},
$1^-1^-$~\cite{a11} and $N\overline{N}$~\cite{ann}. These analyses reveal that
there exists a relative orthogonal phase between the strong and the
electromagnetic amplitudes in $\jpsi$ decays~\cite{dm2exp, mk3exp, a00, lopez,
a11, ann, suzuki}. As to $\psp$, there is also a theoretical argument which
favors the $\pm90^\circ$ phase~\cite{gerard}. Experimentally, some
analyses~\cite{wymppdk,wymphase,besklks1} based on limited $\VP$ and $\PP$ data
indicate that the large phase is compatible with the data. Moreover, some
efforts have been made to extend the phase study to $\pspp$ decay
phenomenologically~\cite{Wang03b,wangp06} and experimentally~\cite{psppklks}.

The great merit of the phase study lies in that it can provide the valuable
clue for the relation between the strong and the electromagnetic interactions.
Now with the upgraded accelerator and detector, BEPCII/BESIII, on May 2009, the
high luminosity of $3\times 10^{32}$cm$^{-2}$s$^{-1}$ had achieved, which is
the highest luminosity in $\tau$-charm energy region which ever existed. The
106 M $\psi^{\prime}$ and 226 M $J/\psi$ events have been collected, even more
colossal data are to be collected in the forthcoming years, which gives a great
opportunity to determine the phase between the strong and the electromagnetic
amplitudes with unprecedented statistical precision.

A favorable way to measure the phase is through the scan experiment
which is the most model-independent approach. However, even with high
luminosity accelerator, the exclusive scan experiment of charmonium decay is
fairly difficult due to low statistics at each energy point. Therefore, the
optimization study for the data taking strategy is of great importance in order
to obtain the most accurate results with the limited luminosity (equivalently
within the limited data taking time).

Without losing generality, we focus on the mode of $\psp$ decays to two
pseudoscalars. Because, as will be shown in the next section, this decay mode
can accommodate a comparatively simple parametrization form which is of great
benefit to extract the relative phase. To get the optimized data taking scheme,
we resort to the sampling simulation technique which is successfully used in
the study of the data taking strategy for a high precision $\tau$ mass
measurement~\cite{taustat, taustat2}. For such kind of method, great many times
of fits should be done, where a large number of calculations need to be
performed to get the theoretically expected observed cross section.
Unfortunately, two nested integrations in this calculation take too
long time to make the actual optimization procedure impractical.

In this paper, we devote to the simplification of calculation for the
observed cross section of $\psp$ decaying to pseudoscalar pair. Some
reasonable assumptions lead us to obtain the analytic expression for the
Initial State Radiative (ISR) corrected cross section. That is to say, we
transform the two-fold integral into a one-fold integral which speeds up the
calculation by one hundred times.

\section{Observed cross section}

The process of $\psp$ decays to Pseudo-scalar and Pseudo-scalar (PP) final
state can be parameterized by merely two amplitudes~\cite{a11, haber}, that is

\begin{equation}
	\begin{array}{lcl}
		A_{\pp} &=& A_{EM}~~, \\ A_{\kk} &=& A_{EM} +
		A_{S},\\ A_{\kskl} &=& A_{S}~~,
	\end{array}
	\label{em}
\end{equation}
where $A_{EM}$ denotes the electromagnetic amplitude and $A_{S}$ the SU(3)
breaking strong amplitude. Here, the G-parity violating channel $\pp$ is
through the electromagnetic process~(the contribution from the
isospin-violating part of QCD is expected to be small~\cite{chernyak82} and is
neglected), $\kskl$ through the SU(3) breaking strong process, and $\kk$
through both. For $\EE$ experiment, the actual amplitudes must include the
contribution of continuum which features the electromagnetic
process~\cite{Wang03,wymppdk,wymphase}:

\begin{equation}
	\begin{array}{lcl}
		A_{\pp} &=& A^c_{EM}+ A_{EM}~~, \\ A_{\kk} &=&
		A^c_{EM}+ A_{EM} + A_{S},\\ A_{\kskl} &=& A_{S}~~,
	\end{array}
	\label{emc}
\end{equation}
where $ A^c_{EM}$ is the amplitude of the continuum contribution. Besides the
common part, $A^c_{EM}$, $A_{EM}$ and $A_{S}$ can be expressed explicitly as

\begin{equation}
	\begin{array}{lcl}
		A^c_{EM} &\propto&{\displaystyle
		\frac{1}{s}}~~,{\rule[-3.5mm]{0mm}{7mm}} \\ A_{EM}
		&\propto&{\displaystyle \frac{1}{s} B(s)}~~, \\ A_{S}
		&\propto&{\displaystyle {\cal C} e^{i \phi} \cdot \frac{1}{s} B(s)
		}~~,
	\end{array}
	\label{ecem}
\end{equation}
where the real parameters $\phi$ and ${\cal C}$ are the relative phase and the
relative strength between the strong and the electromagnetic amplitudes, and
$B(s)$ is defined as~\cite{wymppdk}

\begin{equation}
	B(s)=\frac{3\sqrt{s}\Gamma_{ee}/\alpha}{s-M^2_{\psp}+iM_{\psp}\Gamma_t}~~.
	\label{bexpr}
\end{equation}
Here $\sqrt{s}$ is the center of mass energy, $\alpha$ is the QED fine
structure constant; $M_{\psp}$ and $\Gamma_t$ are the mass and the total width
of $\psp$; $\Gamma_{ee}$ is the partial width to $\EE$.

The Born order cross sections for the three channels read

\begin{equation}
	\begin{array}{ll}
		\sigma_{Born}^{\pp}(s) &= {\displaystyle
		\frac{4\pi\alpha^2}{s^{3/2}}\left[1+2\Re B(s)+|B(s)|^2\right] } \\ &
		\times {\displaystyle \left| {\cal F}_{\pp} (s)\right|^2{\cal P}_{\pp}
		(s)} ~,
	\end{array}
	\label{bnpp}
\end{equation}

\begin{equation}
	\begin{array}{ll}
		\sigma_{Born}^{\kk}(s)&={\displaystyle
		\frac{4\pi\alpha^2}{s^{3/2}}\left[1+2\Re ({C}_{\phi} B(s)) +\left|{
		C}_{\phi} B(s)\right|^2 \right]} \\ & \times {\displaystyle
		\left|{\cal F}_{\kk} (s) \right|^2{\cal P}_{\kk} (s) }~,
	\end{array}
	\label{bnkk}
\end{equation}

\begin{equation}
	\sigma_{Born}^{\kskl}(s)=
	\frac{4\pi\alpha^2}{s^{3/2}} {\cal C}^2 |B(s)|^2 |{\cal F}_{\kskl}
	(s)|^2{\cal P}_{\kskl} (s) ,
	\label{bnkskl}
\end{equation}
where $C_{\phi} = 1 + {\cal C}e^{i\phi}$; ${\cal F}_{f.s.}(s)=f_{f.s.}/s$, with
$f_{f.s.}$ being an energy independent constant, and $f.s.=\pp,\kk,\kskl$;
${\cal P}_{f.s.}(s) = 2q^3_{f.s.}/3s$, with $ q^2_{f.s.} = E^2_{f.s.} -
m^2_{f.s.} = s/4 - m^2_{f.s.}$.

It is obvious that in Eq.~\eref{bnkk}, if ${ C}_{\phi} =1$,
$\sigma^{Born}_{\kk}(s)$ is identical to $\sigma^{Born}_{\pp}(s)$ while if ${
C}_{\phi} ={\cal C}e^{i\phi}$, $\sigma^{Born}_{\kk}(s)$ is identical to
$\sigma^{Born}_{\kskl}(s)$. From the mathematical point of view, the cross
section expression of $\sigma^{Born}_{\kk}(s)$ is more general with the
expressions of $\sigma^{Born}_{\pp}(s)$ and $\sigma^{Born}_{\kskl}(s)$ as its
special cases.  Therefore, in the following study, only tackled is the formula
for $\kk$ final state and $f.s.$ is simply donated as $K$.

In $\EE$ collision, the Born order cross section is modified by the ISR in the
way~\cite{Kuraev85}

\begin{equation}
	\sigma_{r.c.} (s)=\int_{0}^{X_f} dx F(x,s)
	\frac{\sigma_{Born}(s(1-x))}{|1-\Pi (s(1-x))|^2},
	\label{eq_isr}
\end{equation}
where $X_f=1-s'/s$. $F(x,s)$ has been calculated to an accuracy of
0.1\%~\cite{Kuraev85,Altarelli,Berends} and $\Pi(s)$ is the vacuum polarization
factor. In the upper limit of the integration, $\sqrt{s'}$ is the
experimentally required minimum invariant mass of the final particles. In this
work, $X_f=0.15$ is used which corresponds to invariant mass cut of 3.4~GeV/$c^2$.

By convention, $\Gamma_{ee}$ has the QED vacuum polarization in its
definition~\cite{Tsai,Alexander}. Here it is natural to extend this convention
to the partial widths of other pure electromagnetic decays, that is

\begin{equation}
	\Gamma_{K} = 2
	\tilde{\Gamma}_{ee}\left(\frac{q_{K}}{M_{\psp}}\right)^3 \left|{\cal
	F} (M^2_{\psp}) \right|^2~,
	\label{eq_defgf}
\end{equation}
where

$$
\tilde{\Gamma}_{ee} \equiv
\frac{\Gamma_{ee}}{|1 - \Pi (m^2_{\psp})|^2}~,
$$
with vacuum polarization effect included.

The $\EE$ colliders have finite energy resolution which is much wider than the
intrinsic width of $\psp$. Such energy resolution is usually a Gaussian
distribution~\cite{bk:Lee,bk:Wille}:

$$
G(W,W^{\prime})=\frac{1}{\sqrt{2 \pi} \Delta} e^{
-\frac{(W-W^{\prime})^2}{2 {\Delta}^2} },
$$
where $W=\sqrt{s}$ and $\Delta$, a function of the energy, is the standard
deviation of the Gaussian distribution. The experimentally observed cross
section is the radiative corrected cross section folded with the energy
resolution function

\begin{equation}
	\sigma_{obs} (W)=\int \limits_{0}^{\infty}
	dW^{\prime} \sigma_{r.c.} (W^{\prime}) G(W^{\prime},W).
	\label{eqegsprd}
\end{equation}

For briefness, the variables $\tilde{\Gamma}_{ee}$, $M_{\psp}$, and
$\Gamma_{t}$ are respectively written as ${\Gamma}_{ee}$, $M$, and $\Gamma$
hereafter.

\section{Simplification of ISR correction}

In this section, we focus on the simplification of ISR correction of the
observed cross section. In the energy region we concerned (3.67 GeV/$c^2$ --
3.71 GeV/$c^2$), the vacuum polarization factor could be concerned as constant
and absorbed into $\tilde \Gamma_{ee}$ as in Eq.~\eref{eq_defgf}. So we could begin
with this expression:

\begin{equation}
	\sigma_{r.c.}(s) = \int_0^{X_f} dx F(x,s) \sigma_{Born}(s(1-x))~.
	\label{eqisr}
\end{equation}
%where $X_f = 1-s_m/s$, $\sqrt{s_m}$ is the experimentally required minimum
%invariant mass of the final particles. 
$F(x,s)$ is the structure function, which can
be expressed as follows:

\begin{equation}
	\begin{split}
		F(x,s) = x^{t-1}\cdot B_1(t) + x^t \cdot B_2(t) + \\
		x^{t+1}\cdot B_3(t) + O(x^{t+1}t^2)
	\end{split}~,
	\label{eqfxs}
\end{equation}
where

\begin{equation}
	\begin{array}{ll}
		B_1(t) & ={\displaystyle t\cdot \left[ 1 +
		\frac{\alpha}{\pi}(\frac{\pi^2}{3} - \frac{1}{2}) + \frac{3}{4}t +
		t^2\left(\frac{9}{32} - \frac{\pi^2}{12} \right) \right] }~,\\
		B_2(t) &= {\displaystyle-t-\frac{t^2}{4} }~,\\
		B_3(t) &= {\displaystyle \frac{t}{2} - \frac{3}{8}t^2}~,
	\end{array}
	\label{expbs}
\end{equation}
with

\[
\quad t = \frac{2\alpha}{\pi} (\textrm{ln}\frac{s}{m_e^2} -1)~.
\]

Based on Eq.~\eref{bnkk}, the whole expression of the observed cross section is
subdivided into three terms: the continuum, the resonance, and the interference
terms. The simplification of each term will be discussed separately.

\subsection{Continuum term}
\label{sbxn_cterm}

In the light of Eq.~\eref{bnkk}, the Born order expression for the continuum is
written explicitly as

\begin{equation}
	\sigma_{Born}^C = \frac{8\pi\alpha^2
	f_K^2}{3} \cdot \frac{(s/4-m_K^2)^{3/2}}{s^{9/2}}~.
	\label{eqbornc}
\end{equation}

In the above equation, the most crucial part is the factor

\[
l_{9/2}(s)=\frac{(s/4-m_K^2)^{3/2}}{s^{9/2}}~.
\]

\begin{center}
	\includegraphics[angle=-90,width=8cm]{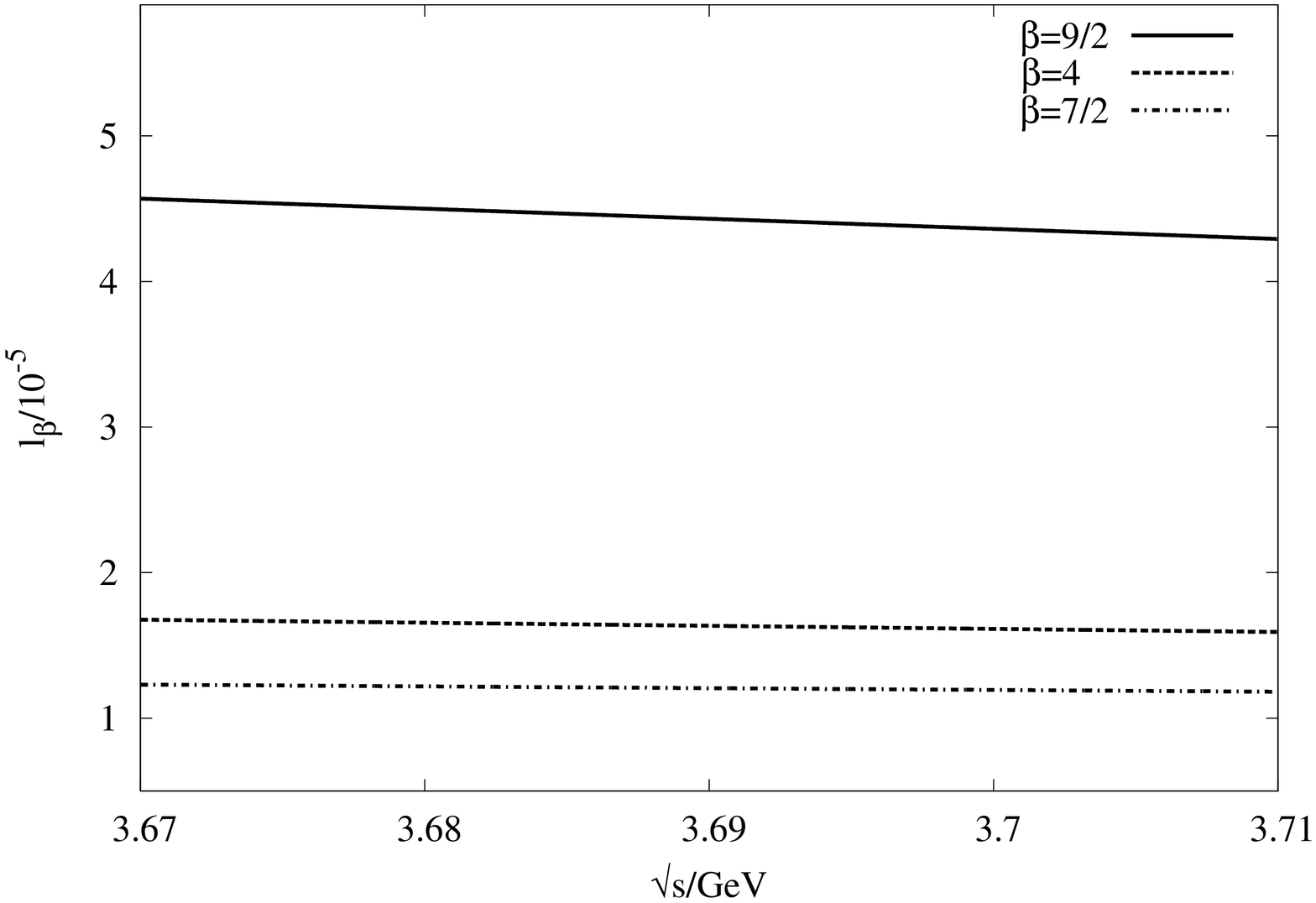}
	\figcaption{\label{figfactor}Variations of factor $l_{\beta}(s)$ against
	center-of-mass energy ($\sqrt{s}$) in the vicinity of $\psp$
	resonance peak for $\beta=9/2,~4,~7/2$. }
\end{center}

For the study of charmonium physics, $s$ is much greater than $m^2_K$,
therefore the factor $l_{9/2}(s)$ variates almost linearly in the vicinity of
$\psp$ peak as shown in Fig.~\ref{figfactor}. With this observation, it is
natural to approximate the factor $l_{9/2}(s)$ with a linear function, viz.

\[
\bar{l}_{9/2}(s) \approx \lambda_{9/2} \cdot s + \zeta_{9/2}~.
\]

As a matter of fact, the similar factors appear in the resonance and
interference terms as well. So generally, we define
\begin{equation}
	l_{\beta}(s)=\frac{(s/4-m_K^2)^{3/2}}{s^{\beta}}~,
\end{equation}
and utilizing the approximation

\begin{equation}
	\bar{l}_{\beta}(s) \approx \lambda_{\beta} \cdot s +
	\zeta_{\beta}~. \label{eqlnrfac}
\end{equation}
Here the coefficients $\lambda_{\beta}$ and $\zeta_{\beta}$ can be determined
analytically, the details are degraded into the appendix~\footnote{The
determination of linear coefficients $\lambda_{\beta}$ and $\zeta_{\beta}$ is
similar to that of linear regression, where the optimization is used. However,
for linear regression, a linear function is used to fit a set of separated data
while for our problem, a linear function is used to approximate another
non-linear function. Such an idea of linearization is referred to as the
generalized linear regression.}.

With the linearization of the factor $l_{9/2}(s)$, the $x$-concerned ISR
integral for the continuum term has actually the form

\begin{equation}
	\rho_0=\int_0^{X_f}x^{\mu} dx~,
	\label{rho0}
\end{equation}
which can be integrated easily. So the ISR corrected cross section of the
continuum is expressed analytically as follows

\begin{equation}
	\sigma_{r.c.}^C =
	\frac{8\pi\alpha^2 f_k^2}{3} \cdot [ (\lambda_{9/2} \cdot s +
	\zeta_{9/2} ) \cdot H_0(s) - \lambda_{9/2} \cdot s \cdot H_1(s)]~,
	\label{xcteqisrc}
\end{equation}
with

\[
H_{\mu}(s) \equiv \int_0^{X_f}x^{\mu}F(x,s)dx =
\sum\limits_{\nu=1}^{3} \frac{X_f^{t+\mu+\nu-1}}{t+\mu + \nu - 1} \cdot
B_{\nu}(t)~.
\]

\subsection{Resonance term}

In the light of Eq.~\eref{bnkk}, the Born order expression for the resonance is
written explicitly as

\begin{equation}
	\sigma_{Born}^R = \frac{8\pi\alpha^2 f_k^2}{3} \cdot \frac{A_1}
	{(s-M^2)^2+M^2\Gamma^2} \frac{(s/4-m_K^2)^{3/2}} {s^{7/2}}~,
	\label{eqbornr}
\end{equation}
where

\[
A_1 = 9\Gamma_{ee}^2/\alpha^2 \cdot (1+{\cal C}^2+2{\cal C}\cos\phi)~.
\]

As far as the factor

\[
l_{7/2}(s)=\frac{(s/4-m_K^2)^{3/2}}{s^{7/2}}~
\]
is concerned, the similar approximation as the previous section is adopted,
viz.

\[
\bar{l}_{7/2}(s) \approx \lambda_{7/2} \cdot s + \zeta_{7/2}~.
\]

The $x$-concerned ISR integral for the resonance term then reads

\begin{equation}
	\rho(s,t) = \int_0^{X_f}\frac{x^{t-1}dx}{(s(1-x)-M^2)^2 +
	M^2\Gamma^2},
	\label{rhost}
\end{equation}
which can be integrated analytically~\cite{wangp90,cahn87}

\begin{equation}
	\begin{aligned}
		\rho(s,t)= \frac{1}{ts^2} \cdot a^{t-2}\frac{\pi t
		\sin[\theta(1-t)]} {\sin\theta \sin\pi t} + \frac{1}{s^2} \cdot
		\left[ \frac{1}{t-2}\cdot X_f^{t-2} + \right. \\
		\frac{2(s-M^2)}{(t-3)s}\cdot X_f^{t-3}
		\left. + \frac{3(s-M^2)^2 - M^2\Gamma^2}{(t-4)s^2}\cdot X_f^{t-4}
		\right]~,
	\end{aligned}
	\label{eqrhost}
\end{equation}
where

\[
a^2 = \left(1-\frac{M^2}{s}\right)^2 + \frac{M^2\Gamma^2}{s^2}
\quad (a>0), \quad \cos\theta = \frac{1}{a} \cdot \left(\frac{M^2}{s}
-1 \right)~.
\]

With the expression of $\rho(s,t)$, the ISR corrected cross section of the
resonance is re-casted as

\begin{equation}
	\sigma_{r.c.}^R =
	\frac{8\pi\alpha^2 f_K^2}{3} \cdot A_1 \cdot [ (\lambda_{7/2} \cdot s
	+ \zeta_{7/2} ) \cdot G_0(s) - \lambda_{7/2} \cdot s \cdot G_1(s)]~,
	\label{xcteqisrr}
\end{equation}
with
\begin{equation}
	\begin{array}{rcl}
		G_{\mu}(s) & = &
		\int_0^{X_f}\frac{x^{\mu}\cdot F(x,s)dx}{(s(1-x)-M^2)^2 + M^2\Gamma^2}\\
		&=& \sum\limits_{\nu=1}^{3} \rho(s,t+\mu+(\nu-1))\cdot
		B_{\nu}(t)~.
	\end{array}
	\label{intgmu}
\end{equation}

\subsection{Interference term}

The Born order expression for the interference can be acquired readily from
Eq.~\eref{bnkk}. However, for clearness the expression of the interference is
further divided into two sub-terms as follows

\begin{equation}
	\sigma_{Born}^{I_1} = \frac{8\pi\alpha^2 f_K^2}{3} \cdot
	\frac{A_2\cdot(s-M^2)}{(s-M^2)^2+M^2\Gamma^2} \cdot
	\frac{(s/4-m_k^2)^{3/2}}{s^{4}}~,
	\label{eqborni1}
\end{equation}
and

\begin{equation}
	\sigma_{Born}^{I_2} = \frac{8\pi\alpha^2 f_k^2}{3} \cdot \frac{A_3}
	{(s-M^2)^2+M^2\Gamma^2} \cdot \frac{(s/4-m_k^2)^{3/2}}{s^{4}}~,
	\label{eqborni2}
\end{equation}
where

\[
A_2 = 6\left(\Gamma_{ee}/\alpha\right) \cdot (1+{\cal C}\cos\phi)~,
\]

\[
A_3 = 6\left(\Gamma_{ee}/\alpha\right) \cdot {\cal C} M\Gamma\sin\phi~.
\]

The simplification strategy is the same as those used for the continuum and
resonance. First, the factor

\[
l_{4}(s)=\frac{(s/4-m_k^2)^{3/2}} {s^{4}}~
\]
is approximated as

\[
\bar{l}_{4}(s) \approx \lambda_{4} \cdot s + \zeta_{4}~;
\]
second, the $x$-concerned ISR integrals for the interference terms have the
forms as those in Eqs.~\eref{rho0} and \eref{rhost}, which can be integrated
out directly or by Formula \eref{eqrhost}. Finally, the ISR corrected cross
section of the interference is obtained

\begin{equation}
	\begin{split}
		\sigma_{r.c.}^{I_1} = \frac{8\pi\alpha^2 f_k^2}{3} \cdot A_2
		\cdot \{(\lambda_{4} \cdot s + \zeta_{4})(s-M^2) \cdot G_0 \\
		- [2\lambda_{4} \cdot s^2 +
		(\zeta_{4} - \lambda_{4} M^2)s ] \cdot
		G_1(s)+ \lambda_{4} s^2 \cdot G_2(s) \}~,
	\end{split}
	\label{xcteqisri1}
\end{equation}

\begin{equation}
	\sigma_{r.c.}^{I_2} = \frac{8\pi\alpha^2 f_k^2}{3} \cdot A_3 \cdot
	[(\lambda_{4} s + \zeta_{4}) \cdot G_0(s) - \lambda_{4} s \cdot
	G_1(s)] ~,
	\label{xcteqisri2}
\end{equation}
where $G_{\mu}(s)$ is given by Formula \eref{intgmu}.

In summary, the ISR corrected cross section formula is
\begin{equation}
	\sigma_{r.c.} (s) = \sigma_{r.c.}^{C} (s)+ \sigma_{r.c.}^{R} (s)+
	\sigma_{r.c.}^{I_1}(s) + \sigma_{r.c.}^{I_2} (s)~,
	\label{eqofsumsim}
\end{equation}
with expressions of the cross section for each term given in
Eqs.~\eref{xcteqisrc}, ~\eref{xcteqisrr}, ~\eref{xcteqisri1}, and
~\eref{xcteqisri2}, respectively.

\section{Possible simplification for energy spread integral}

As indicated in Eq.~\eref{eqegsprd}, the experimentally observed cross section
is the $\sigma_{r.c.}$ convoluted $G(W^{\prime},W)$, which might be simplified
further. Two methods, the Taylor Expansion (TE) method and the Fast Fourier
Transformation (FFT) method, have been considered for such a simplification.

For the TE method, we begin from Eq.~\eref{eqegsprd}, and Taylor expand the
$\sigma_{r.c.}$ at W, viz.

$$
\sigma_{r.c.}(W^{\prime}) = \sum\limits_{n=0}^{\infty}
\frac{\sigma_{r.c.}^{(n)}(W)}{n!} \cdot (W^{\prime}-W)^{n}~,
$$
where $\sigma_{r.c.}^{(n)}(W)$ denotes the $n$-th derivative of function
$\sigma_{r.c.}$ at value $W$.  Replace the Taylor expansion of $\sigma_{r.c.}$
into Eq.~\eref{eqegsprd}, the integral to be calculated has the following form

$$
\int \limits_{-\infty}^{\infty} x^{n} e^{-x^2} dx~,
$$
which can be precalculated. However, in order to achieve a reasonable
precision, we need to calculate hundreds, or even thousands of terms in Taylor
expansion. This means that the fairly high order derivatives of $\sigma_{r.c.}$
have to be calculated,and too much time is consumed, which is not acceptable.

As to FFT method\footnote{http://en.wikipedia.org/wiki/Fast\_Fourier\_transform}, 
we could easily find that the observed cross
section $\sigma_{obs}(W)$ is a convolution of the radiative corrected cross
section and a gauss function. Considering the Convolution Theorem in Fourier
Transformation

$$
\mathfrak{F}(g\bigotimes h) = \mathfrak{F}(g) \cdot \mathfrak{F}(h)~,
$$
where $\mathfrak{F}$ represents Fourier Transformation, $\bigotimes$ represents
convolution. To calculate convolution efficiently, we use Fast Fourier
Transformation. First, $\sigma_{r.c.}$ and $G$ should be sampled in energy
region. After that, we get two series of numbers. Then DFT (Discrete Fourier
Transformation) should be performed on both series, and the resulting series
should be multiplied to generate one final series. Finally IDFT (Inverse
Discrete Fourier Transformation) should be performed on this series and what we
get is the distribution of $\sigma_{obs}$ in energy region on which
$\sigma_{r.c.}$ and $G$ are sampled. This process is very fast, and we could
get the result on the whole energy region at the same time rather than
calculating the integral one by one. To get accurate result, the sample number
should be very large (512 or 1024), which means a large number of cross
sections should be calculated. In real energy scan, the number of data taking
points is usually not large (less than 20). The total integration time in a
small number of energy points is less than the time cost by sampling a large
number of cross sections and perform DFT and IDFT on it. So this method does
not fit our purpose.

\section{Investigation of simplified formula}

\subsection{Precision}

The accurate observe cross section ($\sigma_{obs}$) is calculated by
Eq.~\eref{eqegsprd} while the simplification one (denoted by $\sigma^s_{obs}$)
is also calculated by Eq.~\eref{eqegsprd} but with ($\sigma_{r.c.}$) replaced
by the expression \eref{eqofsumsim}. The relative error of two observed cross
sections is defined as

\begin{equation}
	R_{\sigma} =\frac{\sigma^s_{obs}-\sigma_{obs}}{\sigma_{obs}}~.
	\label{eqrer}
\end{equation}

In the calculation of the observed cross section, all parameters of resonances
are taken from PDG08~\cite{pdg08}, $\Delta=1.3$ MeV is used. Two real
undetermined parameters are the relative phase ($\phi$) and the relative
strength (${\cal C}$) between the strong and the electromagnetic amplitudes.
The dependences of $R_{\sigma}$ on $\phi$ and ${\cal C}$ are shown in
Figs.~\ref{erang} and~\ref{erstg} respectively.

\begin{center}
	\includegraphics[angle=-90,width=8cm]{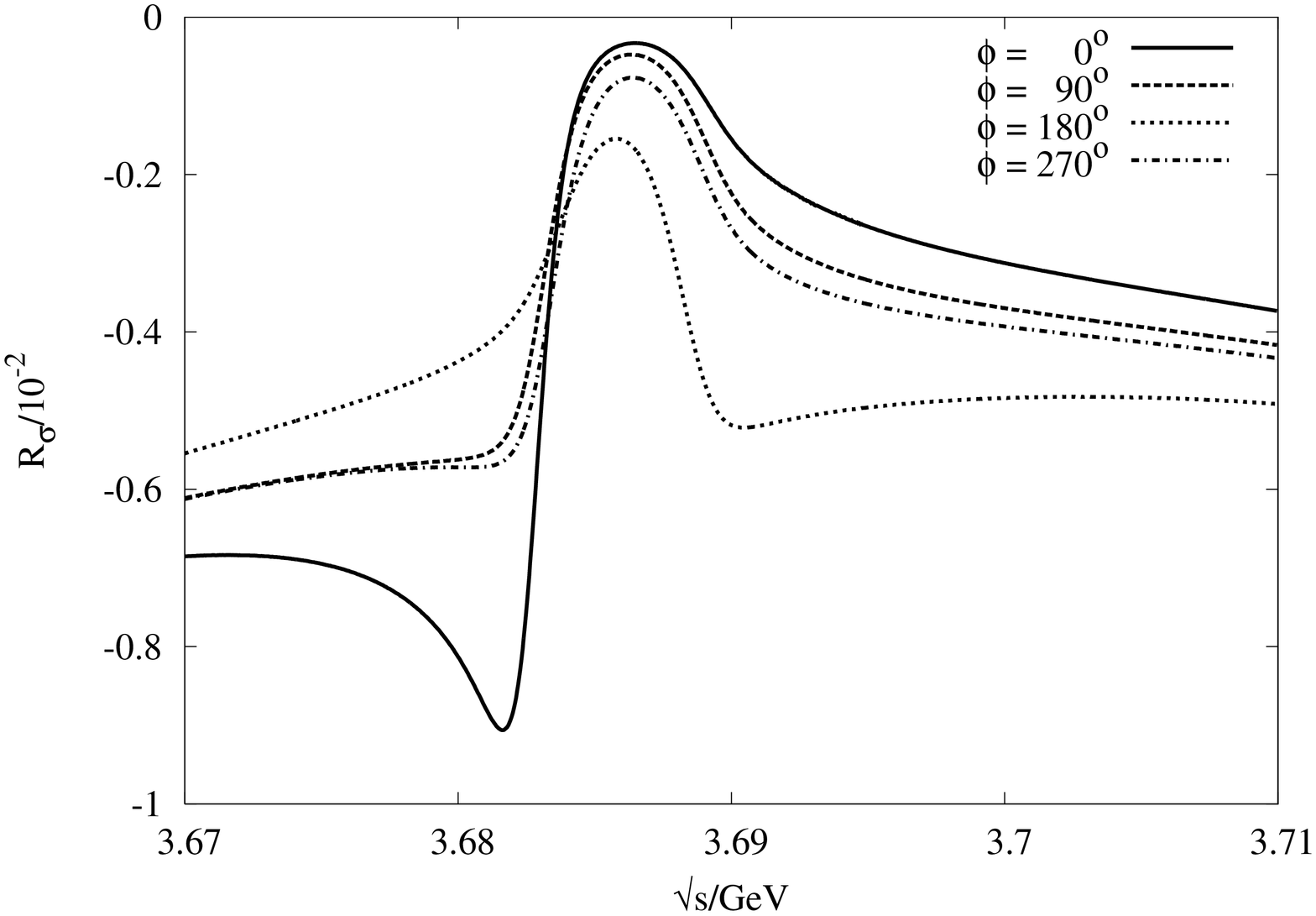}
	\figcaption{\label{erang}Variations of $R_{\sigma}$ against $\sqrt{s}$
	in the vicinity of $\psp$ resonance peak for $\phi=0^{\circ}$,
	$90^{\circ}$, $180^{\circ}$, and $270^{\circ}$. In the calculation
	of the observed cross section, ${\cal C}$ is fixed at 2.5.}
\end{center}

\begin{center}
	\includegraphics[angle=-90,width=8cm]{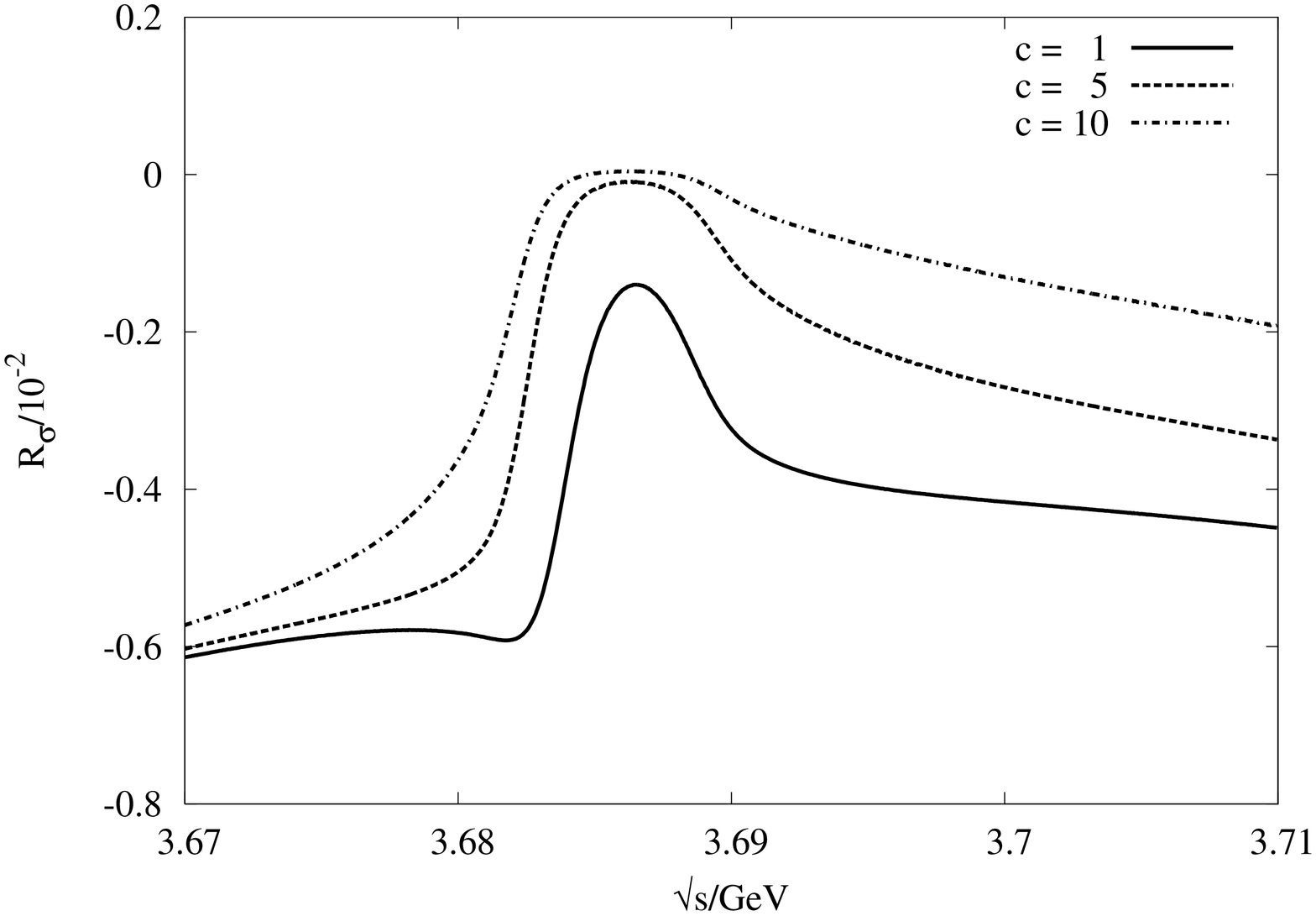}
	\figcaption{\label{erstg}Variations of $R_{\sigma}$ against $\sqrt{s}$
	in the vicinity of $\psp$ resonance peak for ${\cal C}=$1, 5, and
	10. In the calculation of the observed cross section, $\phi$ is
	fixed at $90^{\circ}$.}
\end{center}

The variations of $R_{\sigma}$ against the center-of-mass energy ($\sqrt{s}$)in
the vicinity of $\psp$ resonance peak for $\phi=0^{\circ}$, $90^{\circ}$,
$180^{\circ}$, and $270^{\circ}$ are displayed in Fig.~\ref{erang}, according
to which we notice that firstly, the absolute value of $R_{\sigma}$ is less
than one percent in the energy region we concerned; secondly, the difference
between two cross sections fades away at the resonance peak; thirdly, the
differences in off-resonance region are larger than that in on-resonance
region.  The similar dependence of $R_{\sigma}$ on ${\cal C}$ can be seen from
Fig.~\ref{erstg}, where displayed are the variations of $R_{\sigma}$ against
$\sqrt{s}$ in the vicinity of $\psp$ resonance peak for ${\cal C}=$1, 5, and
10. It is obvious that the difference due to the variation of ${\cal C}$ is
even smaller, which is at the level of a few per mille.

\subsection{Computation time}
The symbol $T^s$ ($T^0$) denotes the computation time when $\sigma^s_{obs}$
($\sigma_{obs}$) is used for the cross section calculation. The comparison of $T^s$
(denoted by the solid line) and $T^0$ (denoted by the dotted line) at both resonance
and off-resonance regions are shown in Fig.~\ref{figtimecp}.

\begin{center}
	\includegraphics[angle=-90,width=8cm]{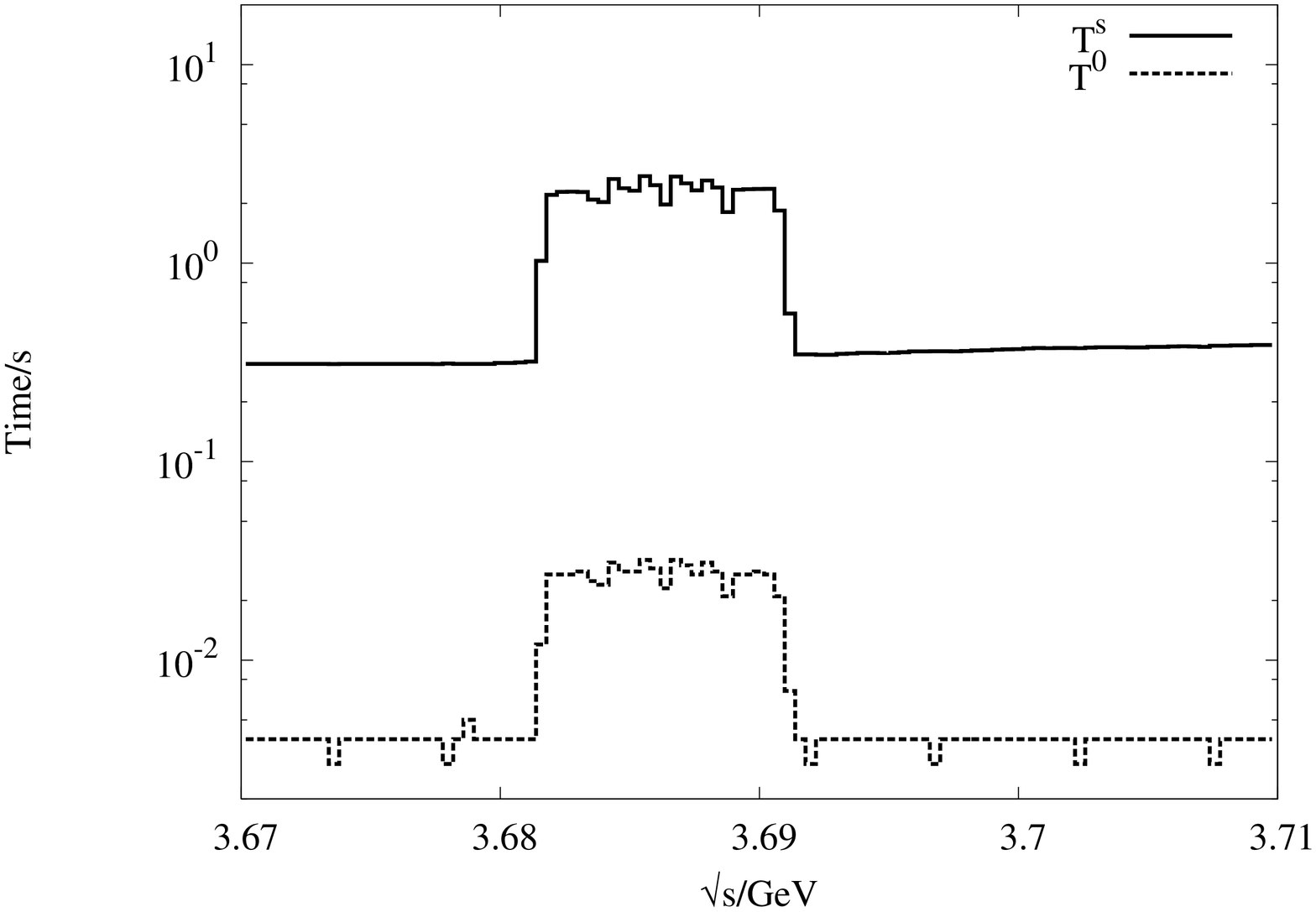}
	\figcaption{\label{figtimecp}Comparison of $T^s$ and $T^0$ at both
	resonance and off-resonance regions.}
\end{center}

From comparison it can be seen that about one-hundred-time reduction of
computation time is achieved by our simplification algorithm.  Although only
one-fold integral is simplified by analytic expression, the computation time is
less than 0.1 second for each energy point which is fast enough for our scan
simulation study.

\subsection{Application}

As we mentioned in the introduction, the speed of calculation of the observed
cross section is the crucial issue of data taking optimization study of the
scan experiment. Without reasonably simplified formula, it will be a too long
time to perform the optimization fit, and the detailed scan optimization is
impractical.

Besides the application in scan optimization, simplified cross section formulas
can also used for the uncertainty study~\cite{moxh01} and correlation
study~\cite{moxhpsdctrep}. Since for both of these studies, the
sampling-and-fitting method is also adopted, the fast computation of cross
section is needed as well.

\section{Summary}

The complete expressions for $\psp \rightarrow PP$ decays are presented, 
including the relative phase between the strong and the electromagnetic
amplitudes. After linearizing one non-linear kinematic factor, the integrand
with the initial state radiation is integrated analytically. Such a
simplification of two-fold integral into a one-fold integral reduces the total
computing time by about one hundred times.

The possible approaches for simplification of energy spread integral are also
discussed.

The simplified formulas of the observed cross sections obtained in this
paper provide a practical tool for the further optimization study of the
scan data taking, which is of great importance for the study of the relative
phase between the strong and the electromagnetic amplitudes.

\end{multicols}
\vspace{10mm}
\begin{multicols}{2}

\subsection*{Appendices A}

\begin{small}

As we have noted in Subsection 3.1, the factor
%~\ref{sbxn_cterm}, the factor

\begin{subequations}
	\renewcommand{\theequation}{A\arabic{equation}}

	\[
	l_{9/2}(s)=\frac{(s/4-m_K^2)^{3/2}}{s^{9/2}}~
	\]
varied almost linearly in the vicinity of $\psp$ peak, and its variation
against $s$ is shown in Fig.~\ref{figfactor}. Therefore, for the factor

\begin{equation}
l_{\beta}(s)=\frac{(s/4-m_K^2)^{3/2}}{s^{\beta}}~,
\end{equation}
a linear function (it refers to Eq.~\eref{eqlnrfac}),

\begin{equation}
\bar{l}_{\beta}(s) \approx \lambda_{\beta} \cdot s + \zeta_{\beta}~
\end{equation}
is utilized to approximate it in the vicinity of resonance peak. The
coefficients $\lambda_{\beta}$ and $\zeta_{\beta}$ are determined by the
generalized linear regression method. As the first step, we define the
integration

\begin{equation}
I=\int_{s1}^{s2}ds[(\lambda_{\beta} \cdot s +
\zeta_{\beta})-\frac{(s/4-m_K^2)^{3/2}}{s^{\beta}}]^2~.
\end{equation}

The needed values of coefficients $\lambda_{\beta}$ and $\zeta_{\beta}$ are
obtained by the minimization of the integration $I$, that is

\begin{equation}
\frac{\partial{I}}{\partial{\lambda_{\beta}}}=0 ~~ \text{and}~~
\frac{\partial I}{\partial \zeta_{\beta}} =0~.
\end{equation}

From the above requirements, we acquire a set of linear equations of
$\lambda_{\beta}$ and $\zeta_{\beta}$, solve it, we obtain

\begin{equation}
\lambda_{\beta}=\frac{\delta_1C_1-\delta_2C_2}{\delta_1\delta_3 -
\delta_2^2}~~ \text{and} ~~\zeta_{\beta} =
\frac{\delta_3C_2-\delta_2C_1}{\delta_1\delta_3 - \delta_2^2} ~,
\end{equation}
where

\[
\delta_i = \int_{s_1}^{s_2} s^{i-1}ds = \frac{s_2^i-s_1^i}{i}~,
\]

\[
C_1 = \int_{s_1}^{s_2}ds\frac{(s/4-m_K^2)^{3/2}}{s^{\beta-1}} =
\frac{1}{8}D(\beta-1)~,
\]

\[
C_2 = \int_{s_1}^{s_2}ds\frac{(s/4-m_K^2)^{3/2}}{s^{\beta}} =
\frac{1}{8}D(\beta)~.
\]

Both $C_1$ and $C_2$ contain integral

\begin{equation}
D(\beta)=\int_{s_1}^{s_2} dx \frac{(x-u)^{3/2}}{x^{\beta}},
\end{equation}
where
\[
\beta= 2,\frac{5}{2},3,\frac{7}{2},4,\frac{9}{2}
\]
\[
u=4m_K^2~.
\]
For different $\beta$, we can calculate the integral analytically~\footnote{The
following integrals are obtained by using Mathematica and checked by hands.}.
For $\beta=2$,

\begin{equation}
D(2)=\left[\sqrt{x-u}\left(\frac{u}{x}+2\right)-3\sqrt{u} \textrm{tan}^{-1}
\left(\frac{\sqrt{x-u}}{\sqrt{u}}\right)\right]\bigg|_{s_1}^{s_2}~;
\end{equation}
For $\beta={\displaystyle \frac{5}{2}}$,

\begin{equation}
D\left(\frac{5}{2}\right)=\left[2\textrm{log}
\left(2\left(\sqrt{x-u}+\sqrt{x}\right)\right) +
\frac{2}{3}\left(\frac{u}{x^{\frac{3}{2}}}-
\frac{4}{\sqrt{x}}\right)\sqrt{x-u}\right]\bigg|_{s_1}^{s_2}~;
\end{equation}
For $\beta=3$,

\begin{equation}
D(3)=\left[\frac{3}{4\sqrt{u}}\textrm{tan}^{-1}
\left(\frac{\sqrt{x-u}}{\sqrt{u}}\right) + \frac{1}{4}\left(\frac{2u}{x^2}-
\frac{5}{x}\right)\sqrt{x-u}\right]\bigg|_{s_1}^{s_2}~;
\end{equation}
For $\beta={\displaystyle \frac{7}{2}}$,

\begin{equation}
D\left(\frac{7}{2}\right)=\frac{2(x-u)^{\frac{5}{2}}}
{5ux^{\frac{5}{2}}}\bigg|_{s_1}^{s_2}~;
\end{equation}
For $\beta=4$,

\begin{equation}
D(4)=\left[\frac{\textrm{tan}^{-1}\left(\frac{\sqrt{x-u}}
{\sqrt{u}}\right)}{8u^{\frac{3}{2}}} + \sqrt{x-u}\left(\frac{u}{3x^3} +
\frac{1}{8ux} - \frac{7}{12x^2}\right)\right]\bigg|_{s_1}^{s_2}~;
\end{equation}
For $\beta={\displaystyle \frac{9}{2}}$,

\begin{equation}
D\left(\frac{9}{2}\right)=\frac{2(x-u)^{\frac{5}{2}}(5u+2x)}
{35u^2x^{\frac{7}{2}}}\bigg|_{s_1}^{s_2}~.
\end{equation}

It could be easily checked that for the coefficients $\lambda_{\beta}$ and
$\zeta_{\beta}$ we obtain,
\begin{equation}
\frac{\partial^2 I}{\partial \lambda_{\beta}^2} =
2\int_{s_1}^{s_2}s^2\textrm{d}s = \frac{2}{3} \left(s_2^3-s_1^3\right)>0~,
\end{equation}
\begin{equation}
\frac{\partial^2 I}{\partial \zeta_{\beta}^2} = 2\int_{s_1}^{s_2}\textrm{d}s =
2(s_2 - s_1)>0~.
\end{equation}
This means what we get is the minimum of $I$, not maximum.

The relative error between the linearized formula and the original formula is
defined as follows:
\begin{equation}
R_{l}=\frac{|\bar{l}_{\beta}-l_{\beta}|}{l_{\beta}}~.
\end{equation}
When $\beta=9/2,~4,~7/2$, the variations of $R_{l}$ against the center-of-mass
energy ($\sqrt{s}$) are shown in Fig.~\ref{betaerror}.

\end{subequations}
\begin{center}
	\includegraphics[angle=-90,width=8cm]{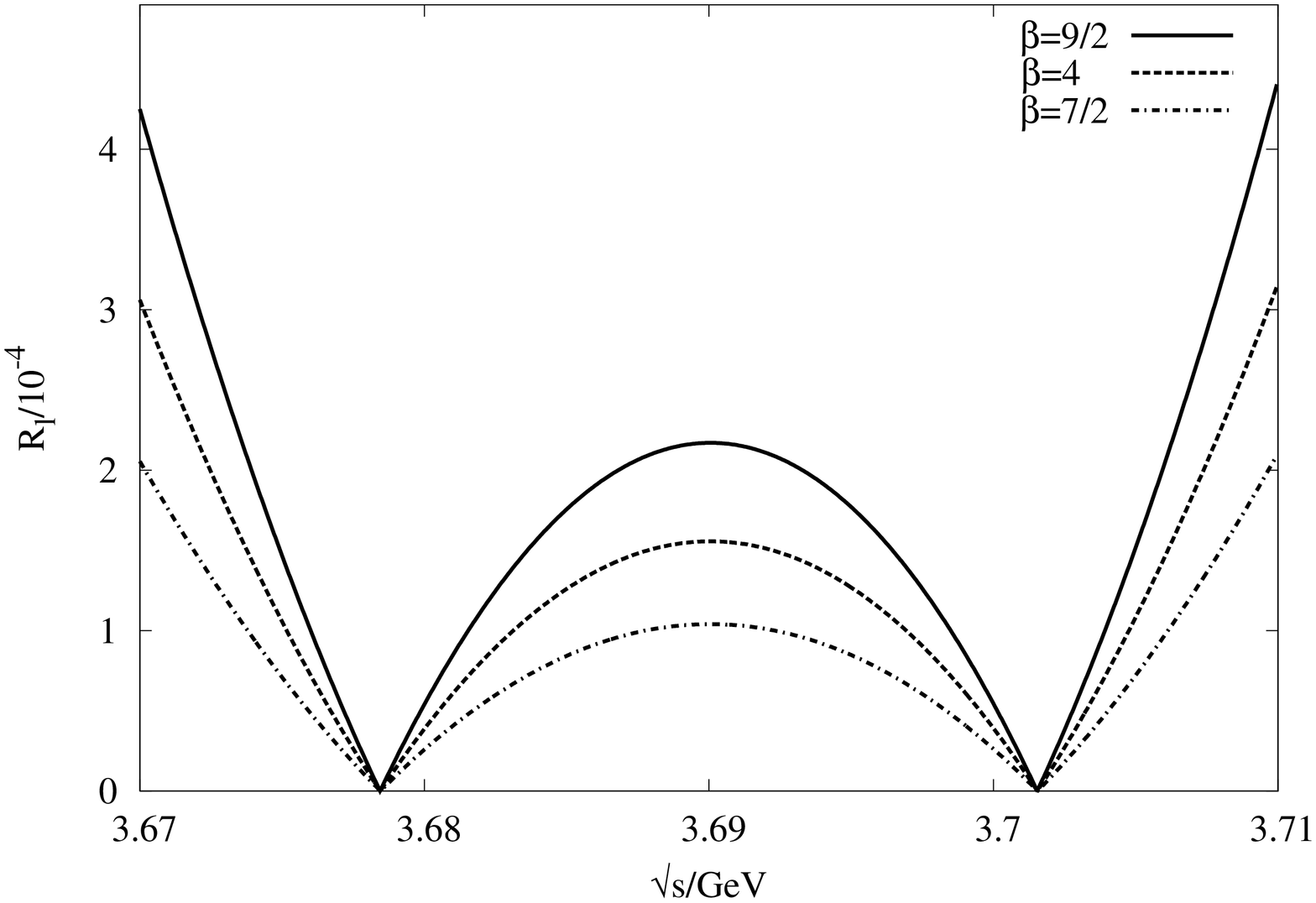}
	\figcaption{\label{betaerror}The variations of $R_{l}$ against
	$\sqrt{s}$ for $\beta=9/2,~4,~7/2$.}
\end{center}

\end{small}
\end{multicols}

\vspace{-1mm}
\centerline{\rule{80mm}{0.1pt}}
\vspace{2mm}

\begin{multicols}{2}

\end{multicols}

\clearpage
\end{document}